\documentclass[twocolumn,english]{revtex4-2}
\usepackage[T1]{fontenc}
\usepackage[latin9]{inputenc}
\usepackage{color}
\usepackage{amsmath}
\usepackage{graphicx}
\usepackage{comment}
\usepackage{graphicx}
\usepackage{sidecap}
\usepackage[colorlinks=true, allcolors=red]{hyperref}

\makeatletter
\usepackage{babel}

\makeatother

\usepackage{babel}
\begin{document}
\title{Nonlinear Dynamics of Current-Carrying ELM Filaments: Spiral Vorticity, Rotation, and Velocity Suppression}

\author{Souvik Mondal$^{1,3,a}$}
\email{$^{a)}$Email: souvik.mondal@ipr.res.in} % Adding the corresponding author's email address
\author{N Bisai$^{1,3}$}
\author{Abhijit Sen$^{1,3}$}
\author{Indranil Bandyopadhyay$^{1,2,3}$}

\affiliation{$^1$Institute for Plasma Research, Bhat, Gandhinagar 382428, Gujarat, India \\
$^2$ITER-India, Institute for Plasma Research, Bhat, Gandhinagar 382428, Gujarat, India \\
$^3$Homi Bhabha National Institute, Training School Complex, Anushaktinagar, Mumbai 40094, India}
%\email{E-mail: souvik.mondal@ipr.res.in\\ souvikmondalphysics421@gmail.com}
%------------------------------------------------------------------------------------------------------------------------------------------------------------------%
\begin{abstract}
%------------------------------------------------------------------------------------------------------------------------------------------------------------------%
In this work, we investigate the nonlinear dynamics of isolated current-carrying edge-localized mode (ELM) filaments using a reduced electromagnetic fluid model in slab geometry. Numerical simulations show that unidirectional parallel current significantly suppresses radial filament velocity and reduces the outward propagation velocity by weakening the curvature-driven interchange force. The reduction in radial velocity is found to follow a modified scaling relation, demonstrating that increasing current progressively weakens outward filament propagation. Analysis of the vorticity equation shows that the electromagnetic current source changes from a dipolar structure to a remarkable spiral pattern, and overcomes the conventional curvature drive in the nonlinear phase. This current-driven source directly imprints its topology on the vorticity field, resulting in spiral vorticity, enhanced angular momentum, increased rotational energy, and localized shear layers. The filament therefore undergoes a transition from a conventional propagating state to a rotationally self-organized electromagnetic structure. These findings demonstrate that parallel current acts as an effective electromagnetic vorticity source and provides new insight into the nonlinear dynamics of ELM filaments in tokamak edge plasmas.

\end{abstract}

\maketitle

% %=========================================================================================================

\section{Introduction}

% %=========================================================================================================

Edge-localized modes (ELMs) are among the most important transient phenomena occurring in high-confinement ($H$-mode) tokamak plasmas. They arise from pressure-gradient and current-driven instabilities at the plasma edge and lead to intermittent expulsions of particles and energy into the scrape-off layer (SOL)~\cite{H_Zohm_1996_ELM, Wang_2013, Kass_1998, DIIID_ELM_PRL, RFX_ELM_PRL, Banerjee_2021_ELM}. The ELM activity, while beneficial for impurity control, can lead to damaging transient heat loads on plasma-facing components for large-amplitude ELM events, which is a major challenge for long pulse operation of future burning plasma devices such as ITER~\cite{Loarte_2007_ELM, PITTS2011S957}. Thus, understanding the nonlinear dynamics of ELM transport is important for predicting and controlling edge plasma behavior.

Experimental and theoretical investigations have demonstrated that ELM transport is achieved by the ejection of coherent filamentary plasma structures, often referred to as ELM filaments or ``blob'' ~\cite{Umansky_1998, KRASHENINNIKOV2001368, Zweben_2002, J_L_Terry_2003, Bisai_2004, bisai_3f2f_2004, Bisai_2005, Shankar_2021}. These filamentary structures are seen to propagate radially across magnetic field lines and are thought to be important in the transient cross-field transport of particles, momentum, and heat in the tokamak edge region~\cite{Garcia2006, Myra2006, KRASHENINNIKOV_2008}. In the conventional picture, the main driver for filament propagation is the curvature-driven interchange mechanism that polarizes the filament and results in a dipolar electrostatic potential structure. This creates a poloidal electric field which drives radial $E\times B$ motion, producing the characteristic mushroom-like deformation frequently observed in experiments and simulations~\cite{KRASHENINNIKOV2001368,dippolito_convective_2011}.

The majority of theoretical studies of blob and filament dynamics have been based on electrostatic or weakly electromagnetic models, where filaments are assumed to be either current-free or with symmetric dipolar current structures~\cite{dippolito_convective_2011}. In these regimes, the evolution of filaments is mainly controlled by interchange forcing, polarization currents, and dissipation mechanisms. However, experimental measurements from several tokamak experiments indicate that ELM filaments can carry considerable field-aligned currents~\cite{Kirk_2006, myra_current_carrying_filament_2007, Current_mes_ELM_PRL, souvik_pop}. Such parallel flows can strongly modify the electromagnetic structure of the filament, possibly affecting its propagation, stability, and nonlinear evolution. The role of the unidirectional parallel current in determining the dynamics of an isolated filament is still not well understood, despite the growing experimental evidence for current-carrying ELM filaments.

Electromagnetic effects have been recently shown to be relevant for the filament dynamics both by theoretical and numerical studies, especially for high-$\beta$ edge plasma conditions when inductive and magnetic perturbations are important~\cite{lee_electromagnetic_2015,lee_electromagnetic_pop,stepanenko_impact_2020}. Contributions from electromagnetic current can change the vorticity generation, the potential structure and influence the radial velocity. Furthermore, it is well-known that strong velocity shear reduces turbulence and cross-field transport in magnetically confined plasmas~\cite{burrell_1997, Boedo_2002, Wagner_2007}. In nonlinear fluid and plasma systems, vorticity can self-organize into coherent spiral or vortex structures~\cite{mcwilliams_vortex_dynamics_1990, Sasaki_2019}. However, most of the previous works have focused on filament acceleration, transport scaling or blob interactions, whereas the possibility of current driven rotational self-organization in isolated ELM filaments has been relatively less addressed. In particular, it remains to be seen how the parallel current alters vorticity generation, and whether it can fundamentally reorganize filament dynamics beyond the conventional interchange-driven propagation.

Motivated by these open questions, in this work, we investigate the nonlinear evolution of a single current-carrying ELM filament using a reduced electromagnetic fluid model in slab geometry. We begin with a filament with unidirectional parallel current, unlike the standard dipolar filaments, and explore how the current amplitude affects filament propagation, vorticity generation, and nonlinear flow organization. Special attention is paid to the role of the electromagnetic current contribution in shaping filament dynamics and its effect on radial velocity.

Our results show that a finite parallel current strongly damps the radial motion of filaments and decreases the strength of the poloidal electric field that drives the $E \times B$ transport. More importantly, we find that the electromagnetic current contribution is the dominant source of vorticity generation and the driver of the formation of a pronounced spiral vorticity structure during the nonlinear phase. This vorticity reorganization is accompanied by an increase of the angular momentum, the rotational energy, and the formation of localized shear layers signalling a transition from conventional interchange-dominated propagation to a rotationally self-organized electromagnetic filament state. This work provides a new insight into the nonlinear dynamics of current-carrying ELM filaments, and may have important implications for understanding transient transport in tokamak edge plasmas.

The remainder of the paper is organized as follows. Section II describes the reduced electromagnetic model and the radial velocity scaling. Section III presents the numerical setup and simulation parameters. Section IV discusses the nonlinear dynamics of current-carrying ELM filaments and the associated velocity suppression. Finally, the discussion and conclusions are presented in Section V.

%=========================================================================================================

\section{Model Equations\label{sec:Model_Equations}}
%=========================================================================================================

To study the nonlinear dynamics of current-carrying ELM filaments, we apply a reduced electromagnetic fluid model based on the Braginskii equations~\cite{braginskii1965transport}. The model is formulated in a three-dimensional slab geometry in Cartesian coordinates, with the equilibrium magnetic field directed along the $z$-axis, and the $x$ and $y$ directions corresponding to the radial and poloidal directions, respectively. This simplified geometry isolates the fundamental electromagnetic processes governing filament evolution while retaining the essential physics relevant to edge-localized mode (ELM) filaments.

Experimental observations suggest that ELM filaments may carry large field-aligned current and therefore can be classified as electromagnetic structures rather than purely electrostatic density perturbations. Previous studies have concentrated primarily on the propagation and transport of filaments, with the effect of parallel current on the structure of filaments, vorticity generation, and nonlinear evolution being relatively less studied. In particular, understanding the way that parallel current changes electric field organization and radial transport is important to characterizing ELM filament dynamics. The primary objective of the present work is therefore to examine how unidirectional parallel current alters the evolution of isolated filaments and drives electromagnetic effects.

In the present study, the cold-ion approximation ($T_i=0$ is adopted, such that ion pressure effects are neglected. Electrons are treated as isothermal, and finite plasma beta, $\beta={8\pi nT_e}/{B^2}$, is retained to include electromagnetic effects through the inductive component of the parallel electric field,
\begin{equation}
E_{\parallel}^{ind}=-\frac{1}{c}\frac{\partial A_{\parallel}}{\partial t},
\end{equation}
where $A_{\parallel}$ is the parallel magnetic vector potential.

The reduced model equations are obtained from particle conservation, quasi-neutrality (current continuity), generalized Ohm's law, and Maxwell's equations~\cite{lee_electromagnetic_2015,stepanenko_impact_2020}. Under these assumptions, the dimensional equations governing the filament dynamics are given by
\begin{equation}
\frac{e\rho_s^2}{T_e}n\frac{d\omega}{dt}=\frac{1}{e}\nabla_{\parallel}J_{\parallel}-\frac{g_i}{\Omega_s}\frac{\partial n}{\partial y},\label{eq:unnormalized_vorticity}
\end{equation}

\begin{equation}
\frac{dn}{dt}=\frac{1}{e}\nabla_{\parallel}J_{\parallel}-\frac{g_i}{\Omega_s}\frac{\partial n}{\partial y},\label{eq:unnormalized_continuity}
\end{equation}

\begin{equation}
-\frac{e}{m_e c}\frac{dA_{\parallel}}{dt}=\frac{e}{m_e}\frac{\partial\phi}{\partial z}-
\frac{T_e}{m_e}\nabla_{\parallel}\ln n+\frac{e}{\sigma_{\parallel}m_e}J_{\parallel},
\label{eq:vector_potential}
\end{equation}
where $n$ is the plasma density, $\phi$ is the electrostatic potential, $A_{\parallel}$ is the parallel magnetic vector potential, and $\omega=\nabla_{\perp}^{2}\phi$ is the vorticity.

The characteristic perpendicular scale is determined by the drift scale, $\rho_s={c_s}/{\Omega_s}$, where $c_s=\sqrt{{T_e}/{m_i}}$ is the ion sound speed and $\Omega_s$ is the ion gyrofrequency. The curvature-driven interchange force is represented by the effective gravity term $g_i={2c_s^2}/{R}$, where $R$ denotes the tokamak major radius.

The parallel current density is defined as
\begin{equation}
J_{\parallel}=ne(V_{i\parallel}-V_{e\parallel}),
\end{equation}
while the parallel electrical conductivity is $\sigma_{\parallel}={1.96n_0e^2}/{m_e\nu_{ei}}$, where the electron-ion collision frequency is given by $\nu_{ei}=2.9\times10^{-6}{n_0\ln\Lambda}/{T_e^{3/2}}$, with Coulomb logarithm $\ln\Lambda \approx 10$.

The total convective derivative is written as
\begin{equation}
\frac{d}{dt}=\frac{\partial}{\partial t}+\frac{c}{B}\hat{b}_0\times\nabla\phi\cdot\nabla,
\end{equation}
while the parallel gradient operator is expressed as
\begin{equation}
\nabla_{\parallel}=\frac{\partial}{\partial z}+\left(\frac{\nabla A_{\parallel}}{B_0}\right)\times\hat{b}_0\cdot\nabla,
\end{equation}
where $\hat{b}_0$ is the unit vector directed along the equilibrium magnetic field.

The relationship between the parallel current and magnetic vector potential follows from Maxwell's equation,
\begin{equation}
J_{\parallel}=-\frac{c}{4\pi}\nabla_{\perp}^{2}A_{\parallel}.
\end{equation}

Since the perpendicular advection time scale of the filament is typically much larger than the electron-ion collision time, electron inertia is neglected~\cite{stepanenko_impact_2020}. To focus on the nonlinear perpendicular dynamics, a reduced formulation is employed by averaging along the magnetic field line while retaining the nonlinear electromagnetic coupling through $A_{\parallel}$. Under this approximation,
\begin{equation}
\nabla_{\parallel}J_{\parallel}=\frac{\partial J_{\parallel}}{\partial z}-[A_{\parallel},J_{\parallel}],
\end{equation}
and
\begin{equation}
\nabla_{\parallel}\ln n=\frac{\partial \ln n}{\partial z}-[A_{\parallel},\ln n].
\end{equation}

The electromagnetic current contribution,
\begin{equation}
\frac{\partial J_{\parallel}}{\partial z}-[A_{\parallel},J_{\parallel}],
\end{equation}
appears explicitly in both the density and vorticity equations and plays an important role in modifying filament transport and nonlinear evolution.

Using the normalizations
\[
\frac{n}{n_0}=\hat{n},
\qquad
t\Omega_s=\hat{t},
\qquad
\frac{x,y}{\rho_s}=(\hat{x},\hat{y}),
\]
\[
\frac{J_{\parallel}}{n_0ec_s}=\hat{J}_{\parallel},
\qquad
\frac{e\phi}{T_{e0}}=\hat{\phi},
\qquad
\frac{A_{\parallel}}{B_0\rho_s}=\hat{A}_{\parallel},
\]
and omitting hats for convenience, the normalized equations become
\begin{equation}
\frac{\partial n}{\partial t}=-[\phi,n]+\frac{\partial J_{\parallel}}{\partial z}-[A_{\parallel},J_{\parallel}]-g\frac{\partial n}{\partial y},\label{eq:norm_density_continuty_eq}
\end{equation}

\begin{equation}
n\frac{\partial \omega}{\partial t}=-[\phi,\nabla_{\perp}^{2}\phi]+\frac{\partial J_{\parallel}}{\partial z}-[A_{\parallel},J_{\parallel}]-g\frac{\partial n}{\partial y},\label{eq:norm_vorticity_eq}
\end{equation}

\begin{equation}
\frac{\partial A_{\parallel}}{\partial t}=\frac{\partial}{\partial z}(\ln n-\phi)-[A_{\parallel},\ln n]+\eta \nabla_{\perp}^{2} A_{\parallel}, \label{eq:norm_vector_pot_eq}
\end{equation}
together with
\begin{equation}
J_{\parallel}=-a\nabla_{\perp}^{2}A_{\parallel}, \label{eq:norm_maxwell_eq}
\end{equation}
where $g=\rho_s/R$ is the normalized curvature parameter, $\eta={1}/{\Omega_s\tau_s}$ is the normalized magnetic diffusion coefficient, $\tau_s={4\pi\sigma_{\parallel}\rho_s^2}/{c^2}$ is the magnetic screening time, and $a$ controls the strength of the electromagnetic current response.

%=========================================================================================
\subsection*{Modified Radial Velocity Scaling in the Presence of Parallel Current}
%=========================================================================================

When parallel current effects are included, the normalized vorticity equation can be written as
\begin{equation}
n\frac{d\omega}{dt} = \nabla_{\parallel} J_{\parallel} - g\,\partial_y n,
\end{equation}
where $\omega = \nabla_\perp^2 \phi$ is the vorticity, $\phi$ is the electrostatic potential, $n$ is the density perturbation, and $J_{\parallel}$ is the parallel current.

We consider a filament of perpendicular size $L_\perp$ with density perturbation amplitude $\delta n$ and characteristic current magnitude $J_{\parallel} \sim J_0$. The vorticity is estimated as
\begin{equation}
\omega \sim \frac{\phi}{L_\perp^2}.
\end{equation}

The time derivative of vorticity is approximated as
\begin{equation}
\frac{d\omega}{dt} \sim \frac{\omega}{\tau} \sim \frac{\phi}{L_\perp^2 \tau},
\end{equation}
where $\tau$ is the characteristic evolution timescale.

The curvature and current terms scale as
\begin{equation}
g\,\partial_y n \sim g \frac{\delta n}{L_\perp}, \qquad
\nabla_{\parallel} J_{\parallel} \sim \frac{J_0}{L_\parallel}.
\end{equation}

Balancing the dominant terms gives
\begin{equation}
\frac{\delta n \,\phi}{L_\perp^2 \tau} \sim \frac{J_0}{L_\perp} -g \frac{\delta n}{L_\perp} .
\end{equation}

Using the $\mathbf{E}\times\mathbf{B}$ velocity scaling
\begin{equation}
V_x \sim -\frac{\phi}{L_\perp},
\end{equation}
we obtain $\phi \sim -V_x L_\perp$. Substituting this into the above expression yields
\begin{equation}
\frac{\delta n \,V_x }{L_\perp \tau} \sim \frac{g\,\delta n}{L_\perp} - \frac{J_0}{L_\parallel}.
\end{equation}

The characteristic timescale is taken as the propagation time across the filament size,
\begin{equation}
\tau \sim \frac{L_\perp}{V_x}.
\end{equation}

Substituting this into the previous equation gives
\begin{equation}
{V_x^2} \sim g\,L_\perp  - \frac{L_\perp^2}{\delta n \, L_\parallel}J_0.
\end{equation}

Thus, the modified radial velocity scaling becomes
\begin{equation}
V_x^2 \sim L_\perp^2 \left(\frac{g}{L_\perp} - \frac{J_0}{\delta n \, L_\parallel }\right)
\end{equation}
where $\delta n \sim n_b$ is the amplitude of the density perturbation and $L_\perp \sim \delta$ is the characteristic perpendicular size of the filament. This expression shows that the parallel current reduces the effective curvature drive responsible for radial propagation. As the current amplitude $J_0$ increases, the radial velocity decreases, leading to weaker outward transport. For the simulation parameters considered in the present work, the curvature drive term, $g/L_\perp$, always exceeds the electromagnetic current contribution, $J_0/(\delta n L_\parallel)$, ensuring that the quantity inside the square root remains positive and the scaling remains physically meaningful. As $J_0$ approaches the curvature drive, the radial propagation is significantly suppressed.

%==========================================================================================================%

\section{Numerical Simulation}
\label{sec:Numerical_simulation}

%==========================================================================================================%
The numerical simulations of Eqs.~(\ref{eq:norm_density_continuty_eq})--(\ref{eq:norm_maxwell_eq}) are carried out using plasma parameters representative of high-$\beta$ tokamak edge conditions, summarized in Table~\ref{table:1}. These parameters correspond to plasma conditions near the last closed flux surface (LCFS) and are consistent with those commonly employed in electromagnetic filament studies~\cite{lee_electromagnetic_pop}. Unless otherwise specified, all physical quantities are expressed in normalized units.

\begin{table}%[h!]
\centering
\setlength{\tabcolsep}{12pt}
\begin{tabular}{l c c} 
 \hline
 Parameter & Value & Unit \\ [0.5ex] 
 \hline\hline
 $n_0$ & $1\times10^{14}$ & cm$^{-3}$ \\ 
 $T_e$ & 200 & eV \\
 $B_0$ & $5.3\times 10^4$ & G \\
 $R$ & 600 & cm \\
 $L_\parallel$ & $10^4$ & cm \\
 $c_s$ & $9.98\times10^{6}$ & cm/s \\ 
 $\Omega_s$ & $2.64\times10^{8}$ & s$^{-1}$ \\
 $\rho_s$ & $3.78\times10^{-2}$ & cm \\
 $\nu_{ei}$ & $1.02\times10^{6}$ & s$^{-1}$ \\
 $\sigma_{\|}$ & $4.82\times10^{16}$ & s$^{-1}$ \\
 $g_i$ & $3.3\times10^{11}$ & cm/s$^{2}$\\
 $g$ & $1.25\times10^{-4}$ & normalized \\
 $\delta$ & 0.7 & cm \\ [1ex] 
 \hline
\end{tabular}
\caption{Plasma parameters used in the simulations, representative of high-$\beta$ tokamak edge conditions near the LCFS~\cite{lee_electromagnetic_pop}.}
\label{table:1}
\end{table}

To investigate the nonlinear dynamics of both dipolar and current-carrying ELM filaments, a single filament is initialized in the tokamak edge region. The computational domain sizes are chosen as $L_x=L_y=256\rho_s$, and $L_z=256384\rho_s$, corresponding to the radial, poloidal, and toroidal directions, respectively.

At the initial time ($t=0$), the filament density is prescribed as a Gaussian perturbation centered at $(x_0,y_0)$ in the perpendicular plane,
\begin{equation}
n(\mathbf{r},0)=1+n_b\exp\left[-\frac{(x-x_0)^2+(y-y_0)^2}{\delta^2}\right],
\end{equation}
where $x_0={L_x}/{2}$ and $y_0={L_y}/{2}$.

Here, $n_b$ denotes the amplitude of the density perturbation and $\delta$ represents the characteristic filament width, selected according to the stable filament regime considered in the present study.

To explore the role of parallel current on filament dynamics, an equilibrium field-aligned current is imposed initially in the form
\begin{equation}
J_{\parallel}(\mathbf{r},0)=J_0\exp\left[-\frac{(x-x_0)^2+(y-y_0)^2}{\delta^2}\right],
\end{equation}
where $J_0$ controls the magnitude of the initial parallel current. The case $J_0=0$ corresponds to the conventional dipolar filament, whereas finite values of  $J_0$ describe current-carrying ELM filaments.

The initial parallel magnetic vector potential $A_{\parallel}$ is determined numerically from Eq.~(\ref{eq:norm_maxwell_eq}) through inversion of the Laplacian operator. To isolate the self-consistent generation of nonlinear flows, the electrostatic potential $\phi$ and vorticity $\nabla_{\perp}^{2}\phi$ are initially set to zero. Boundary conditions are chosen to minimize numerical effects at the domain boundaries. Periodic boundary conditions are imposed in the poloidal ($y$) direction, while Neumann boundary conditions are applied in the radial ($x$) and toroidal ($z$) directions for all evolved quantities, including $n$, $\phi$, $\omega$, $J_{\parallel}$, and $A_{\parallel}$.

The reduced electromagnetic equations are solved numerically using the BOUT++ framework~\cite{DUDSON_bout++}. Spatial derivatives in the radial and toroidal directions are computed using fourth-order central finite-difference schemes, whereas Fourier spectral methods are employed in the poloidal direction. Nonlinear advection terms are evaluated using a third-order weighted essentially non-oscillatory (WENO) scheme to accurately capture steep gradients while maintaining numerical stability. Time integration is performed using the CVODE solver.

Unless otherwise stated, all simulations are carried out using a grid resolution of
\begin{equation}N_x \times N_y \times N_z=260\times256\times32,
\end{equation}
which provides satisfactory numerical convergence and good conservation of total energy. The corresponding grid spacings are
\begin{equation}
dx=dy=\rho_s, \qquad dz=8012\rho_s.
\end{equation}

%==========================================================================================================%

\section{Simulation results}

%==========================================================================================================%

In this section, we present numerical simulations of ELM-like current-carrying filaments using the normalized three-dimensional fluid model described in Sec.~\ref{sec:Model_Equations}. The main objective is to investigate how the current-carrying ELM filament dynamics differ from the conventional filament (dipolar blob) and to study the effect of the magnitude of the current on the filament propagation carried by the filament. Unless otherwise stated, all physical quantities shown in the figures are expressed in normalized units, following the normalization procedure described in Sec.~\ref{sec:Model_Equations}.

%=========================================================================================
\subsection{Modification of Filament Dynamics by Parallel Current}
%=========================================================================================

\begin{figure*}
    \centering
    \includegraphics[width=0.99\linewidth]{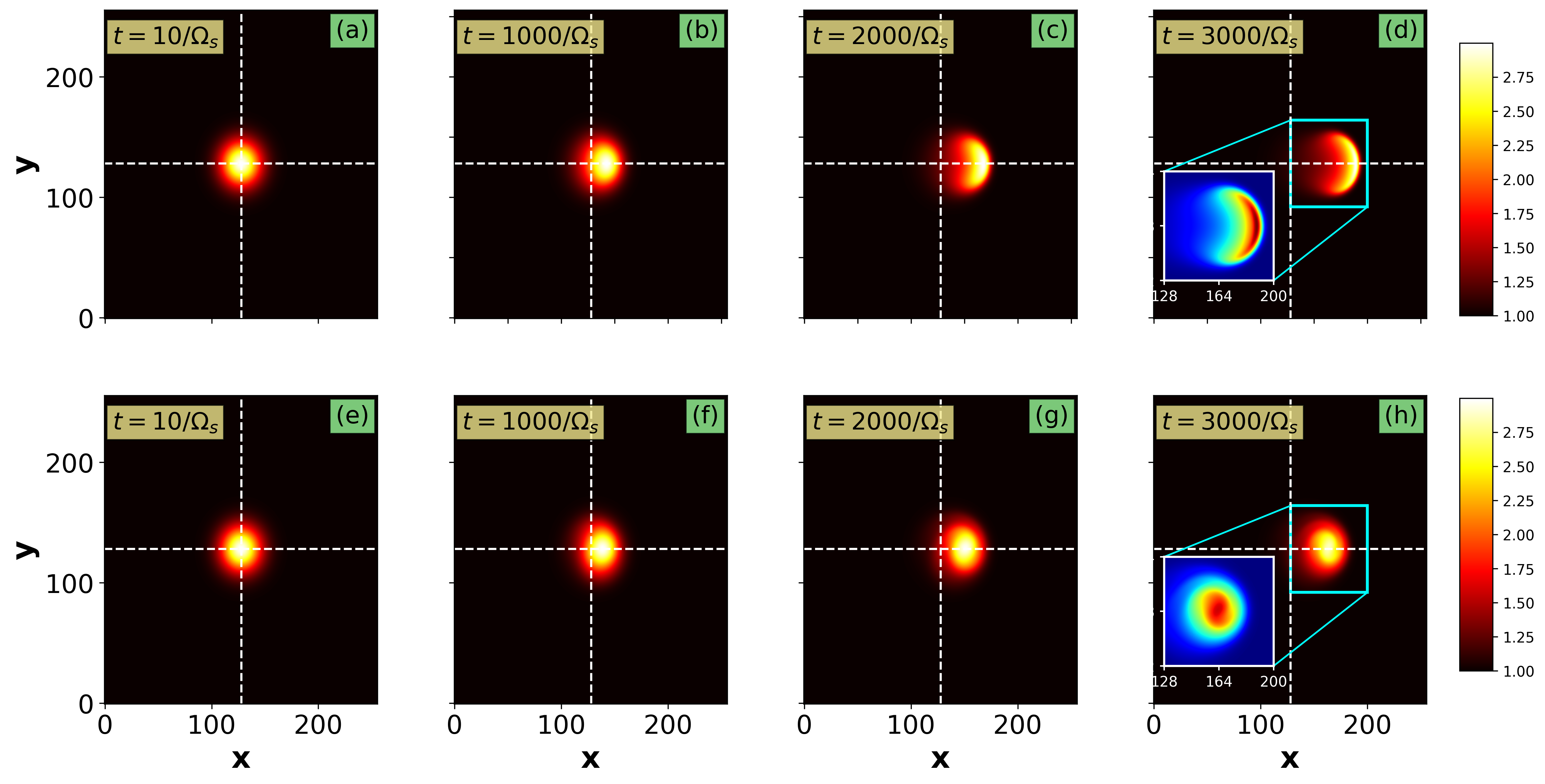}
    \caption{Temporal evolution of the density for the current-free ($J_0=0)$ and current-carrying ($J_0=1.0$) ELM filaments at different times. The dipolar filament develops the characteristic mushroom-like structure and propagates radially outward, whereas the current-carrying filament remains more localized and preserves a compact structure. The inset panels show a magnified view of the filament position at $t=3000/\Omega_s$. The presence of parallel current suppresses radial propagation and modifies the nonlinear filament evolution.}
    \label{fig:density_both_single}
\end{figure*}

We first examine differences in the dynamics between the ELM filaments that carry unidirectional parallel current and the filament with no initial parallel current (dipolar filament or conventional filament). Figure~\ref{fig:density_both_single} shows the evolution of the dipolar or the current-free ($J_0=0$) and current-carrying ELM filament ($J_0=1$). Initially, both filaments have identical density perturbations with magnitude $n_b=2.0$ and are located in the poloidal midplane in the edge region of the tokamak. At the early stage ($t=1-1000 /\Omega_s$), their dynamics are similar and maintain almost a coherent shape. As time increases, in the nonlinear phase ($t=2000-3000 /\Omega_s$)for the dipolar filament ($J_0=0$), the filament propagates radially outward and develops the mushroom-like structure associated with interchange-driven blob dynamics \cite{KRASHENINNIKOV2001368, Garcia2006, dippolito_convective_2011}. The radial displacement increases continuously with time, indicating enhanced outward velocity. On the other hand, in the case of filament carrying finite current remains much more localized and preserves a more compact structure. The radial displacement is visibly smaller, and the filament experiences less deformation. These observations suggest that in the presence of unidirectional parallel current, the nonlinear evolution of the filament modifies and suppresses its outward radial propagation.

\begin{figure*}
    \centering
    \includegraphics[width=0.45\linewidth]{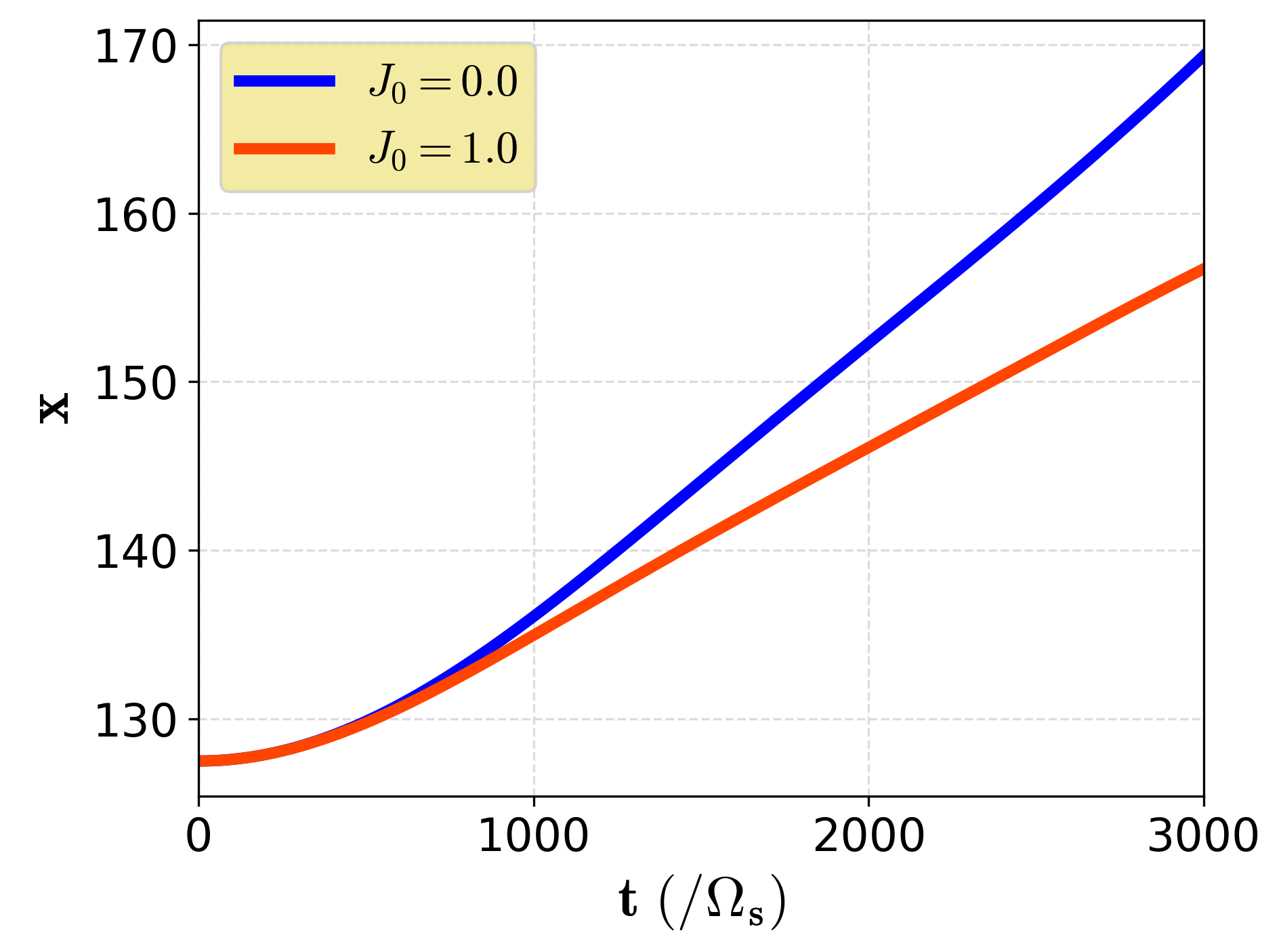}
    \includegraphics[width=0.45\linewidth]{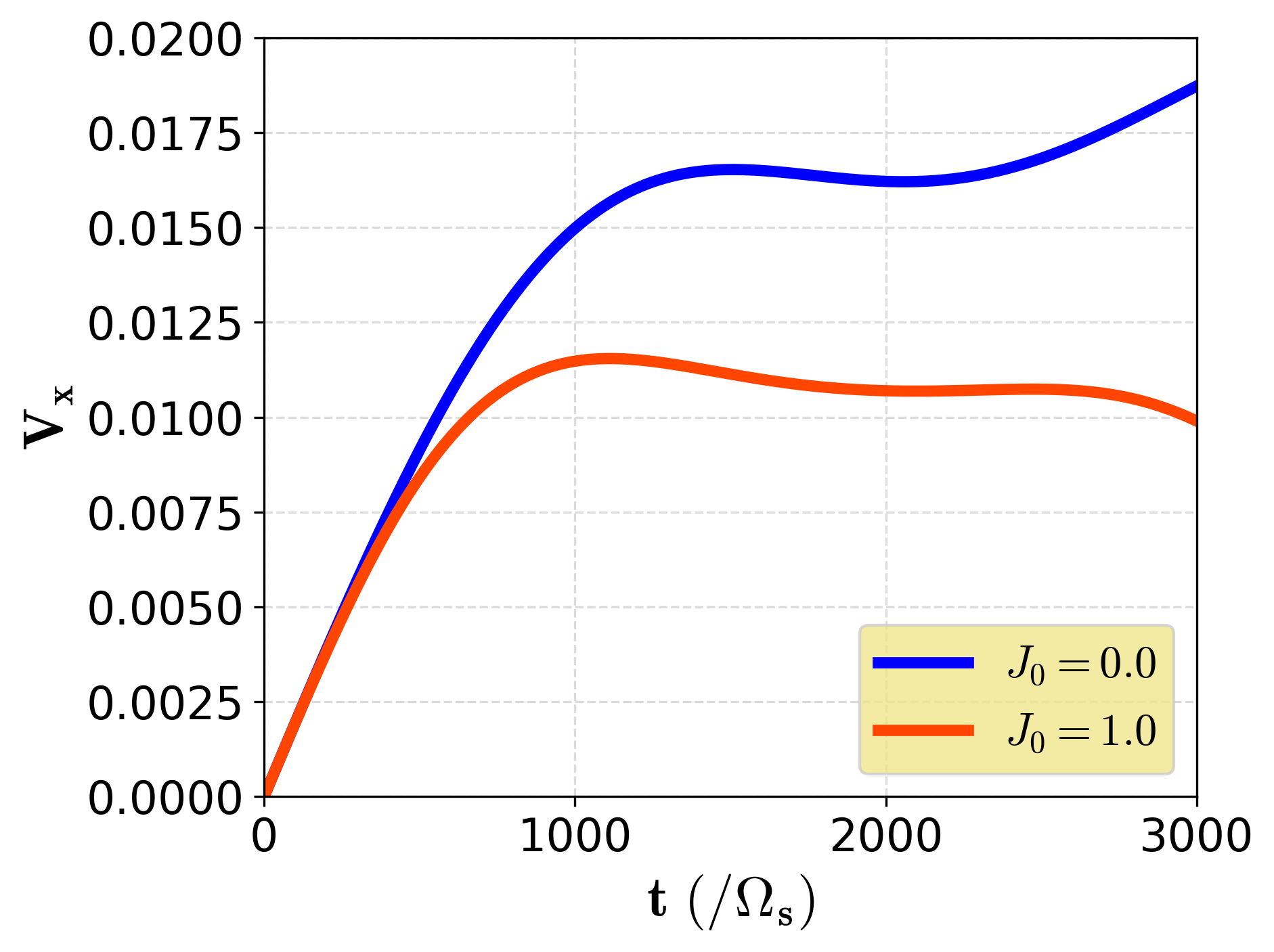}
    \caption{Time evolution of the radial centre-of-mass position ($X_{cm}$) (left) and the average radial velocity (right) for the current-free and current-carrying filaments with different current amplitudes. The current-free filament exhibits stronger outward acceleration, while increasing parallel current progressively reduces both the radial displacement and propagation speed. The suppression becomes more pronounced at higher current amplitudes.}
    \label{fig:radial_velocity}
\end{figure*}

To evaluate the effect of parallel current on filament propagation, we calculate the radial position ($X_{cm}$) and the radial velocity ($V_x$) from the center-of-mass (COM) motion of the density perturbation. The radial position of the filament is calculated from the center of mass of the density distribution as
\begin{equation}
X_{\mathrm{cm}}=\frac{\int x\,(n-n_0)\,dx\,dy}{\int (n-n_0)\,dx\,dy},
\end{equation}
where $n_0$ is the equilibrium background density. The corresponding radial velocity is then obtained from the time derivative of the radial center-of-mass position,
\begin{equation}V_x=\frac{dR_{\mathrm{cm}}}{dt}.
\end{equation}

Figure~\ref{fig:radial_velocity} shows the time evolution of the radial center-of-mass position ($R_{cm}$) (left) and the corresponding radial velocity ($V_x$) (right), for the current-free and current-carrying filaments. During the initial stage of the evolution $\sim (1-800\Omega_s)$, both filaments exhibit nearly identical radial displacement and acceleration, indicating that their dynamics are initially governed by the curvature-driven interchange force. As the nonlinear phase develops, however, the two cases evolve differently. The current-free filament continues to propagate radially outward, resulting in a steady increase in both the center-of-mass position and the radial velocity. In contrast, the current-carrying filament exhibits a much smaller radial displacement, while its radial velocity saturates at a lower value and gradually decreases at later times. Consequently, the filament remains more localized throughout the evolution. These results demonstrate that the presence of parallel current progressively suppresses the outward radial transport of the filament by limiting both its propagation distance and propagation speed.

\begin{figure}
    \centering
    \includegraphics[width=0.99\linewidth]{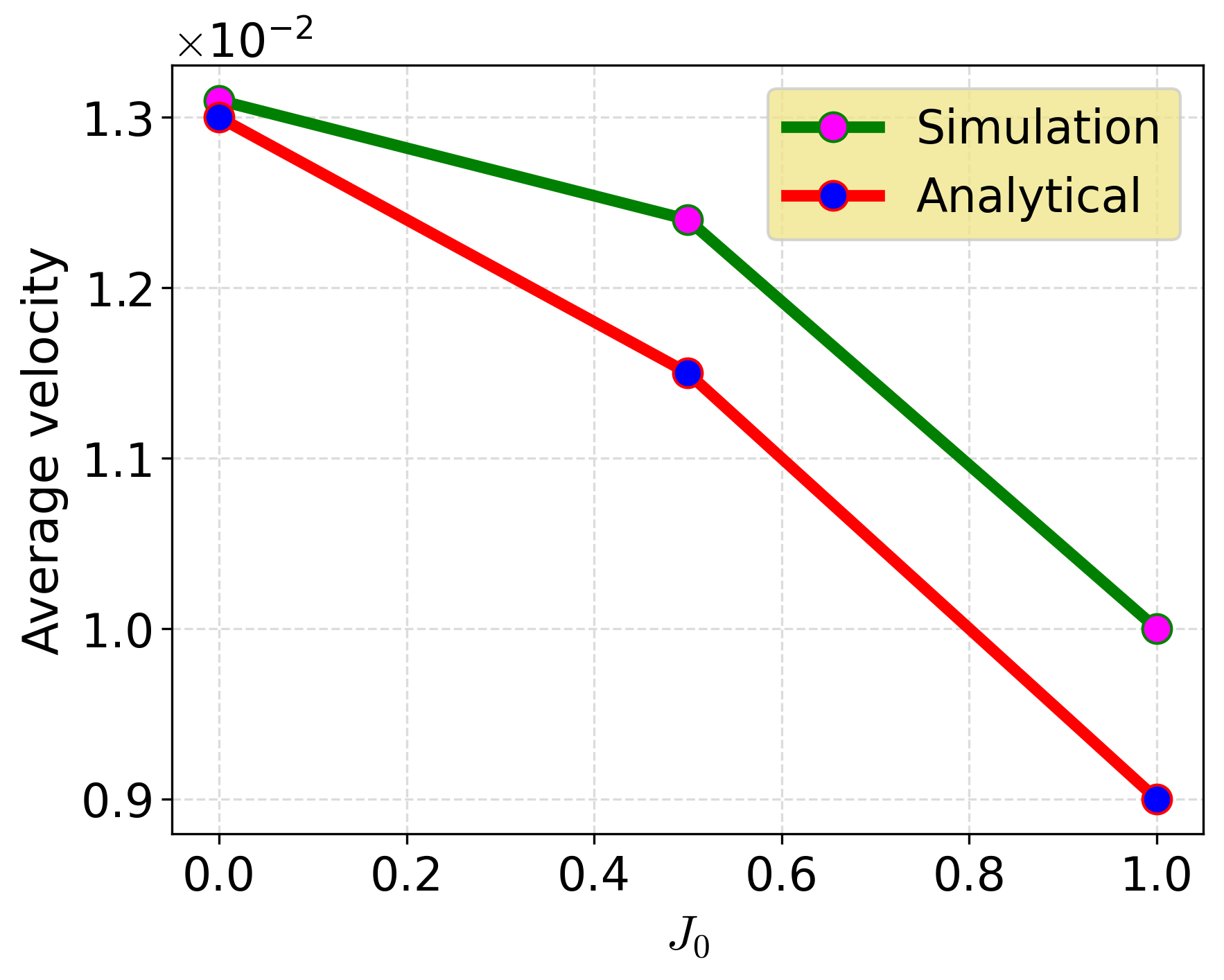}  
    \caption{Comparison of the time-averaged radial velocity obtained from numerical simulations (green) and the analytical scaling (red) as a function of the initial parallel current amplitude, $J_0$. Both the simulation and analytical prediction show a monotonic decrease in the radial velocity with increasing parallel current, demonstrating that stronger current progressively suppresses the outward propagation of the filament. The close agreement between the simulation results and the analytical scaling validates the modified radial velocity model derived in Sec.~\ref{sec:Model_Equations}.}
    \label{fig:avg_velocity_current}
\end{figure}

The dependence of the time-averaged radial velocity on the initial parallel current amplitude is shown in Fig.~\ref{fig:avg_velocity_current}, together with the analytical scaling derived in Sec.~\ref{sec:Model_Equations}. The simulation results show a clear monotonic decrease in the radial velocity with increasing parallel current, indicating that the presence of unidirectional parallel current systematically suppresses the outward radial dynamics of the filament. For lower current amplitudes, the reduction in radial velocity is relatively small, whereas a much stronger suppression is observed for larger current amplitudes. The analytical scaling captures the same trend and agrees well with the simulation results over the entire current range.

This behavior can be understood from the modified radial velocity scaling derived in Sec.~\ref{sec:Model_Equations},
\begin{equation}
V_x \sim L_\perp \sqrt{\left(\frac{g}{L_\perp}-\frac{J_0}{\delta n \, L_\parallel}\right)}.
\end{equation}
The scaling relation shows that the parallel current effectively opposes the curvature-driven interchange force responsible for radial filament propagation. In the absence of parallel current ($J_0=0$), the curvature drive dominates, leading to faster outward motion, consistent with the classical blob propagation mechanism~\cite{Myra2006,dippolito_convective_2011}. As the current amplitude increases, the effective driving force is progressively reduced, resulting in a corresponding decrease in the radial velocity. Consequently, the filament remains more localized and becomes less efficient at transporting plasma across the magnetic field. The close agreement between the analytical prediction and the simulation results shown in Fig.~\ref{fig:avg_velocity_current} validates the proposed scaling and confirms that the electromagnetic current contribution plays an increasingly important role in the nonlinear dynamics of the filament at higher current amplitudes.

%=========================================================================================
\subsection{Current-Driven Vorticity Generation and Rotational Self-Organization}
%=========================================================================================

% \begin{figure*}
%     \centering
%     \includegraphics[width=0.45\linewidth]{data/current_source.png}
%     \includegraphics[width=0.45\linewidth]{data/source_compare_all.png}\hfill
%     \includegraphics[width=0.45\linewidth]{data/omega_delp_jp.png}
%     %\includegraphics[width=0.45\linewidth]{data/avg_velocity_all.png}  
%     \caption{Time evolution of the radial velocity of the current-carrying filaments for both cases (left plot) and the poloidal electric field ($E_y$) in the right plot.}
%     \label{fig:separation}
% \end{figure*}

\begin{figure}
    \centering
    \includegraphics[width=0.99\linewidth]{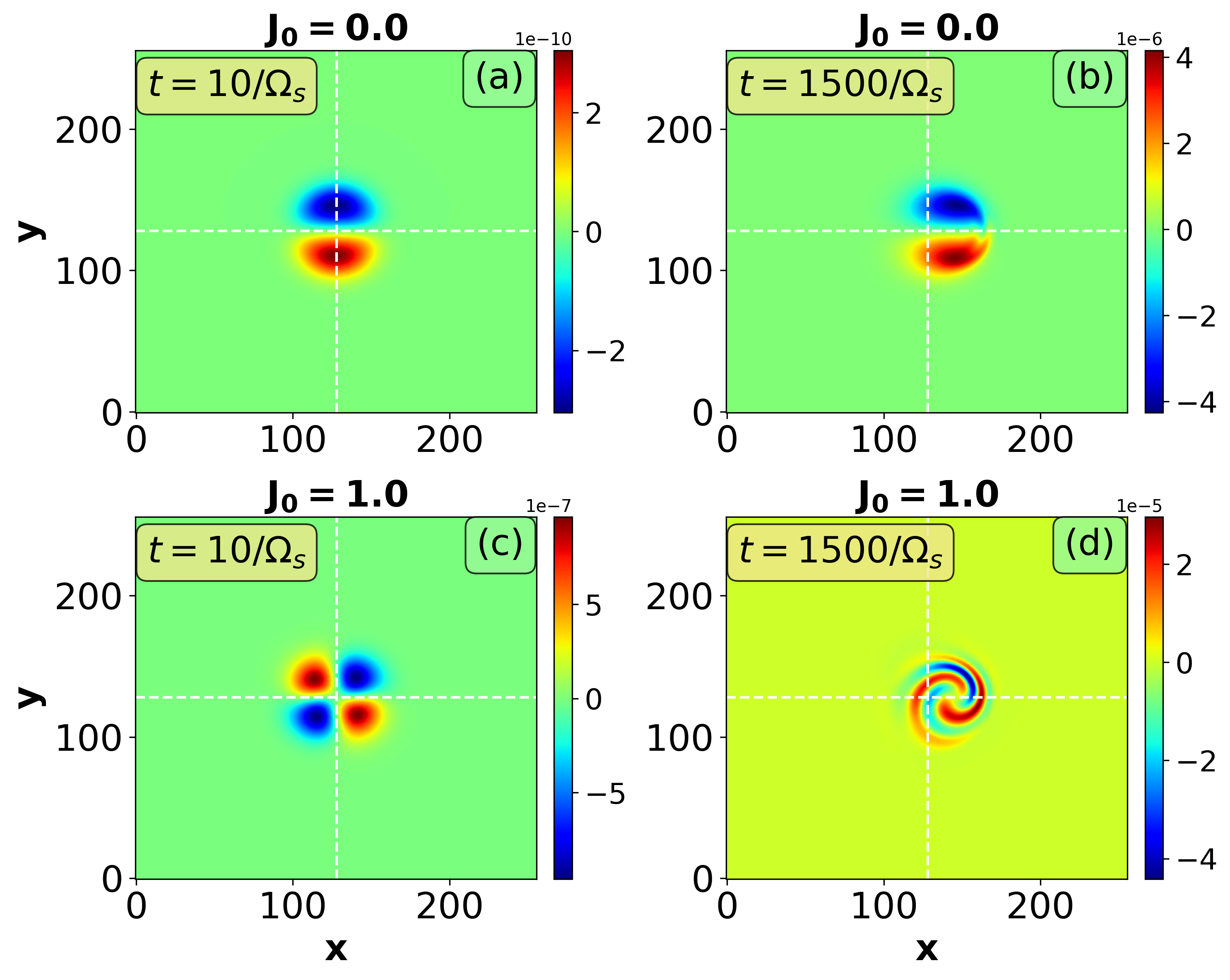}
    \caption{Evolution of the electromagnetic current-driven source term $S_J=\nabla_\parallel J_\parallel$ for the current-free ($J_0=0$) and current-carrying ($J_0=1$) filaments at the initial stage and nonlinear phase ($t=1500/\Omega_s$). While the source term retains an approximately dipolar structure for the current-free filament, the current-carrying filament develops a pronounced spiral topology during nonlinear evolution, indicating strong electromagnetic forcing.}
    \label{fig:current_source}
\end{figure}

\begin{figure}
    \centering
    \includegraphics[width=0.99\linewidth]{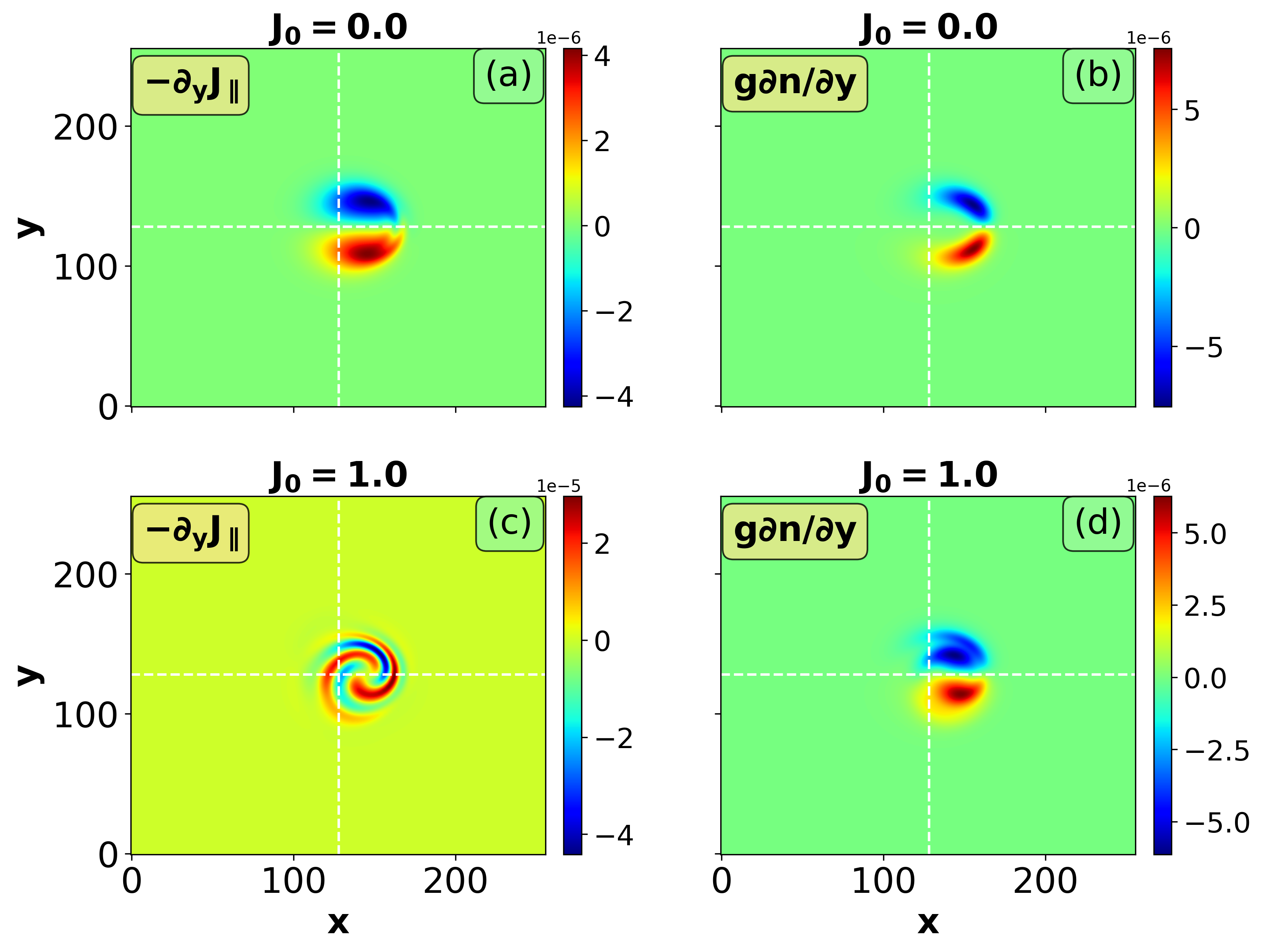} 
    \caption{Comparison between the electromagnetic current contribution $\nabla_\parallel J_\parallel$ and the curvature-driven source term $g\,\partial n/\partial y$ during the nonlinear phase ($t=1500/\Omega_s$) for the current-free and current-carrying filaments. For the dipolar filament, both contributions remain comparable in magnitude and topology. In contrast, the current-carrying filament exhibits a dominant electromagnetic source with a spiral structure, exceeding the conventional curvature drive.}
    \label{fig:source_compare}
\end{figure}

\begin{figure}
    \centering
    \includegraphics[width=0.99\linewidth]{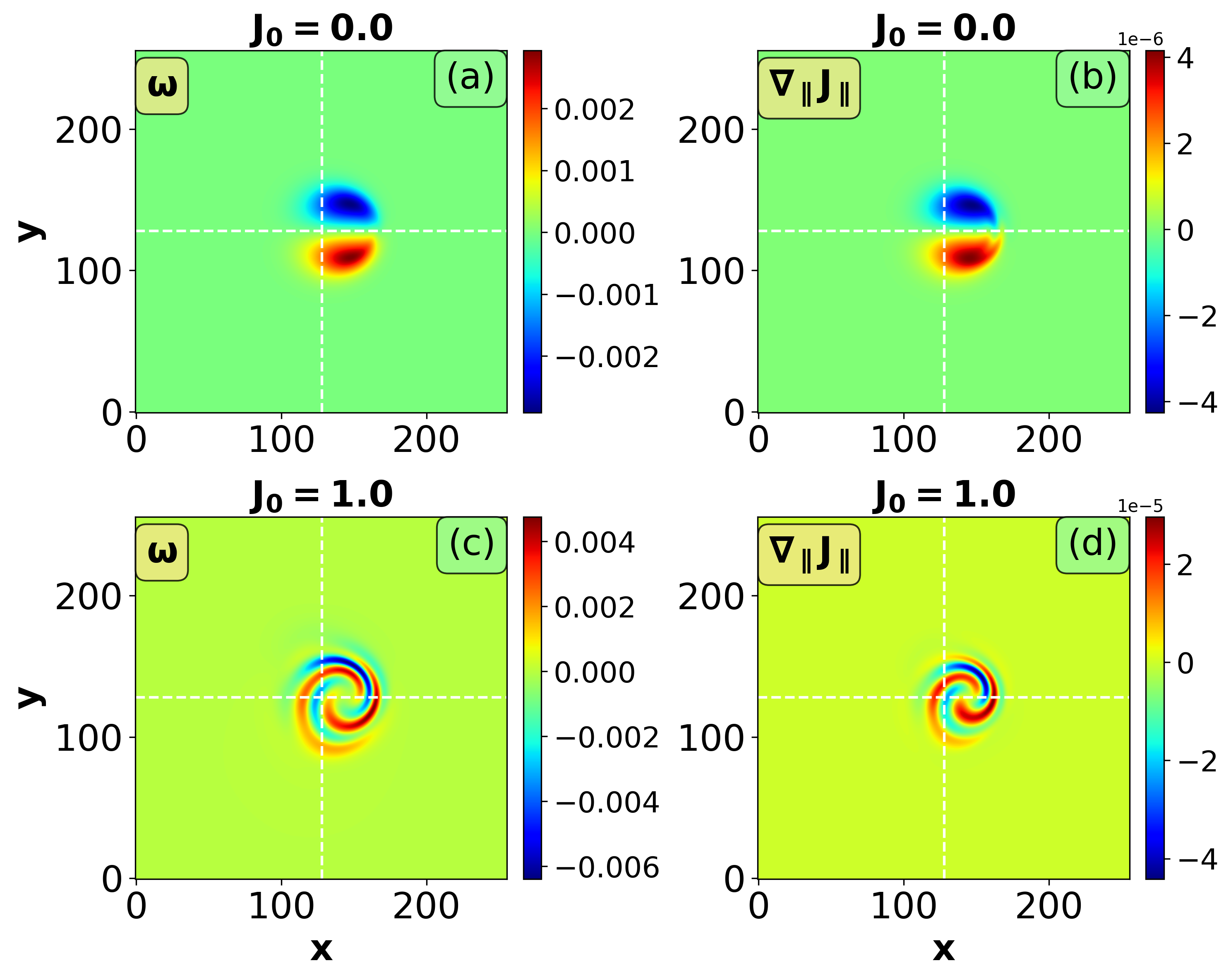}
    \caption{Comparison of the vorticity $\omega$ and the electromagnetic source term $\nabla_\parallel J_\parallel$ during the nonlinear stage ($t=1500/\Omega_s$) for the current-free and current-carrying filaments. The close similarity of the two quantities in the case of a current-carrying system suggests that the contribution of the electromagnetic current directly imprints its topology on the time-evolving vorticity field, leading to the formation of spiral vorticity.}
    \label{fig:omega_source}
\end{figure}

To understand the physical origin of the suppressed radial velocity, we next examine the electromagnetic contribution to the vorticity equation,
\begin{equation}
S_J=\nabla_\parallel J_\parallel=\frac{\partial J_\parallel}{\partial z}-[A_\parallel,J_\parallel],
\end{equation}
which acts as a source term for vorticity generation. 
Figures~\ref{fig:current_source} show the temporal evolution of the current-driven source term for both the current-free ($J_0=0.0$) (a-b) and current-carrying ($J=1$)(c-d) filaments at the initial stage ($t=0$) and during the nonlinear phase of evolution ($t=1500/\Omega_S$). In the absence of parallel current, the source term maintains an approximately dipolar structure throughout the evolution (Fig.~\ref{fig:current_source}a). Even at later times ($t=1500/\Omega_s$), only weak nonlinear deformation appears (Fig.~\ref{fig:current_source}b), indicating that the filament dynamics remain primarily governed by conventional interchange physics. On the other hand, the current-carrying filament undergoes a qualitatively different evolution. Although the source term is initially localized near the filament center with a quadrupole shape (Fig.~\ref{fig:current_source}c), its structure changes significantly as nonlinear effects develop. At later times ($t=1500/\Omega_s$), the current-driven source starts to twist and reorganize into a clear spiral pattern focused around the filament centre as shown in Fig.~\ref{fig:current_source}d. The presence of a spiral structure directly in $S_J$ indicates that the parallel current fundamentally alters the mechanism of vorticity generation and that it introduces a strong rotational forcing which is absent in conventional dipolar filaments. The appearance of spiral vorticity signals the emergence of a coherent self-organized rotational state, similar to vortex organization in nonlinear fluid systems~\cite{Mcwilliams_1984}, but here the organization is driven by electromagnetic current effects.

To further identify the dominant source of vorticity generation, Fig.~\ref{fig:source_compare} (a-d) compares the electromagnetic current contribution, $\nabla_\parallel J_\parallel$, with the conventional curvature-driven term, $g\,{\partial n}/{\partial y}$, during the nonlinear stage of evolution (at $t=1500/\Omega_s$). For the current-free filament, both terms (shown in Fig.~\ref{fig:source_compare} (a-b))exhibit comparable amplitudes and maintain a similar dipolar topology, indicating that the filament evolution remains largely controlled by interchange dynamics. However, a markedly different behavior is observed for the current-carrying filament (shown in Fig.~\ref{fig:source_compare} (c-d)). While the curvature term preserves its conventional dipolar structure, the current-driven contribution develops a pronounced spiral topology and becomes significantly stronger in magnitude. In fact, the electromagnetic source term exceeds the curvature contribution by nearly an order of magnitude, demonstrating that the vorticity dynamics are no longer dominated by curvature effects. Instead, the electromagnetic current contribution becomes the primary mechanism driving the nonlinear evolution.

The role of the current-driven source becomes even more evident in Fig.~\ref{fig:omega_source}(a-d), where the vorticity field $\omega$ is compared directly with the source term $S_J$. For the current-free filament (shown in Fig.~\ref{fig:omega_source}(a-b)), both quantities ($\nabla_\parallel J_\parallel$ and $g\,{\partial n}/{\partial y}$) preserve a conventional dipolar structure, consistent with standard interchange-driven blob motion. In contrast, the current-carrying filament exhibits a striking correspondence between the two fields. The spiral topology observed in the source term is directly reflected in the vorticity distribution (as shown in Fig.~\ref{fig:omega_source}(a-b)), with both structures displaying nearly identical localization, rotational symmetry, and winding direction. Thus, the close correspondence between the current-driven source term and the vorticity field, together with the dominance of $\nabla_{\parallel}J_{\parallel}$ over the curvature drive, demonstrates that parallel current acts as the dominant electromagnetic source of vorticity generation during the nonlinear evolution of the filament. Rather than simply modifying the radial propagation speed, the current fundamentally reorganizes the internal filament dynamics by continuously generating and redistributing vorticity. Consequently, the filament undergoes a transition from a conventional dipolar state to a rotationally self-organized spiral vorticity structure controlled primarily by electromagnetic forcing.

%======================================================================================

%\subsection{Growth of Angular Momentum}

\begin{figure}
    \centering
    \includegraphics[width=0.99\linewidth]{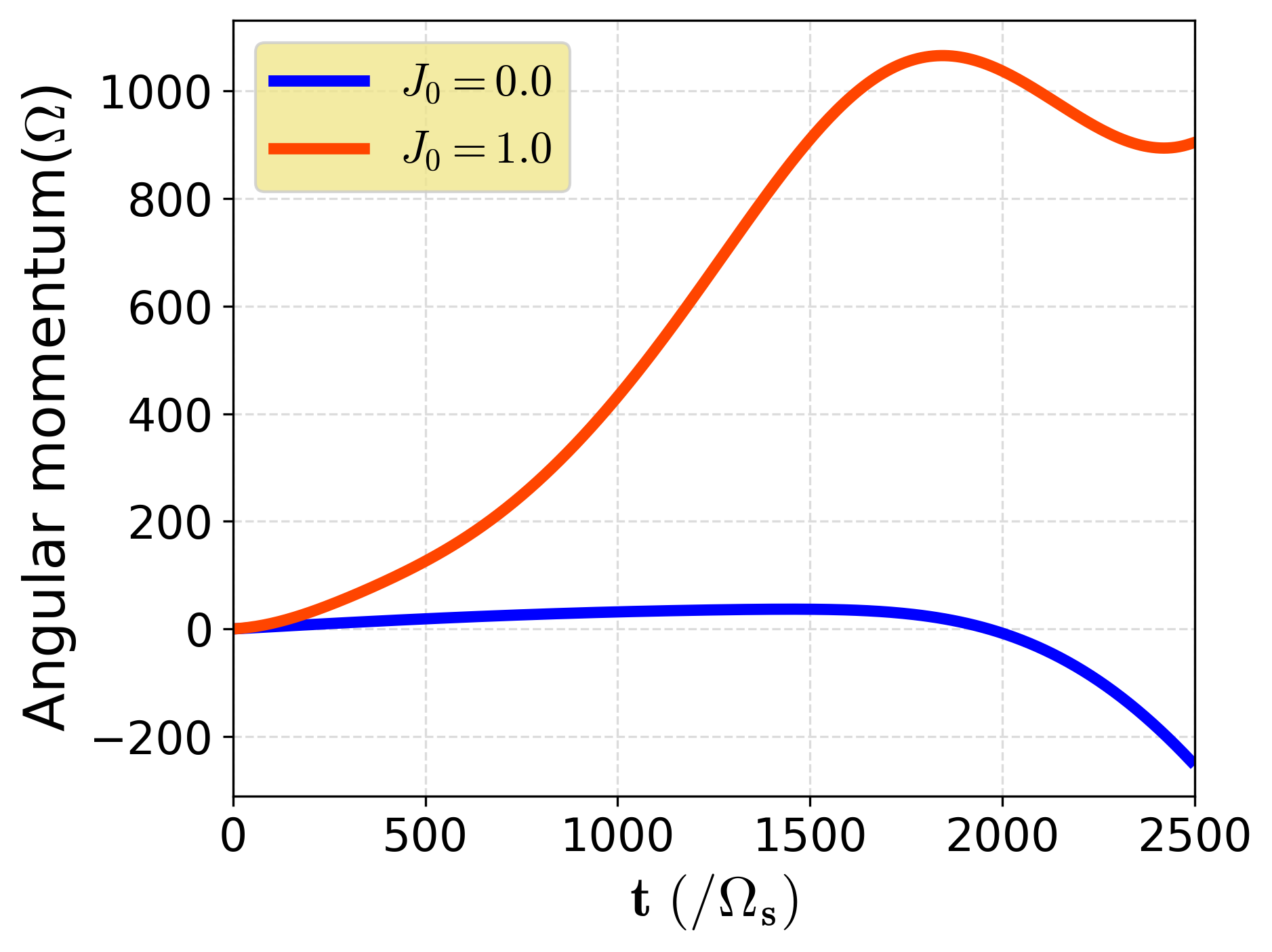}
    \caption{Time evolution of the filament angular momentum for different current amplitudes. The parallel current significantly enhances the angular momentum, which indicates the increased rotational motion and the development of rotational self-organization in the current-carrying filaments.}
    \label{fig:angular_momentum}
\end{figure}

To quantify the rotational dynamics associated with the spiral vorticity state, we evaluate the filament angular momentum as a function of time, shown in Fig.~\ref{fig:angular_momentum}. The angular momentum is calculated from the $E\times B$ velocity field using
\begin{equation}
L=\int\left(x\,v_y-y\,v_x\right)\ dx\ dy,
\end{equation}
where $v_x=-\partial\phi/\partial y$ and $v_y=\partial\phi/\partial x$ represent the normalized radial and poloidal $E\times B$ velocities, respectively.

Figure~\ref{fig:angular_momentum} compares the temporal evolution of angular momentum for the current-free and current-carrying filaments. For the dipolar filament, the angular momentum remains relatively weak throughout the evolution, indicating that the dynamics are dominated primarily by radial propagation. A very different behavior is observed in the presence of parallel current. The angular momentum increases rapidly during the nonlinear phase (from $t=0$ to $t=1700/\Omega_S$) and becomes substantially larger than in the current-free case. The strong increase in angular momentum occurs simultaneously with the appearance of spiral vorticity, indicating a direct connection between vorticity reorganization and rotational motion. Thus, the current-driven electromagnetic source of vorticity plays a key role in generating the enhanced rotational dynamics inside the filament.

%\subsection{Energy Redistribution to Rotational Motion}

\begin{figure}
    \centering
    \includegraphics[width=0.99\linewidth]{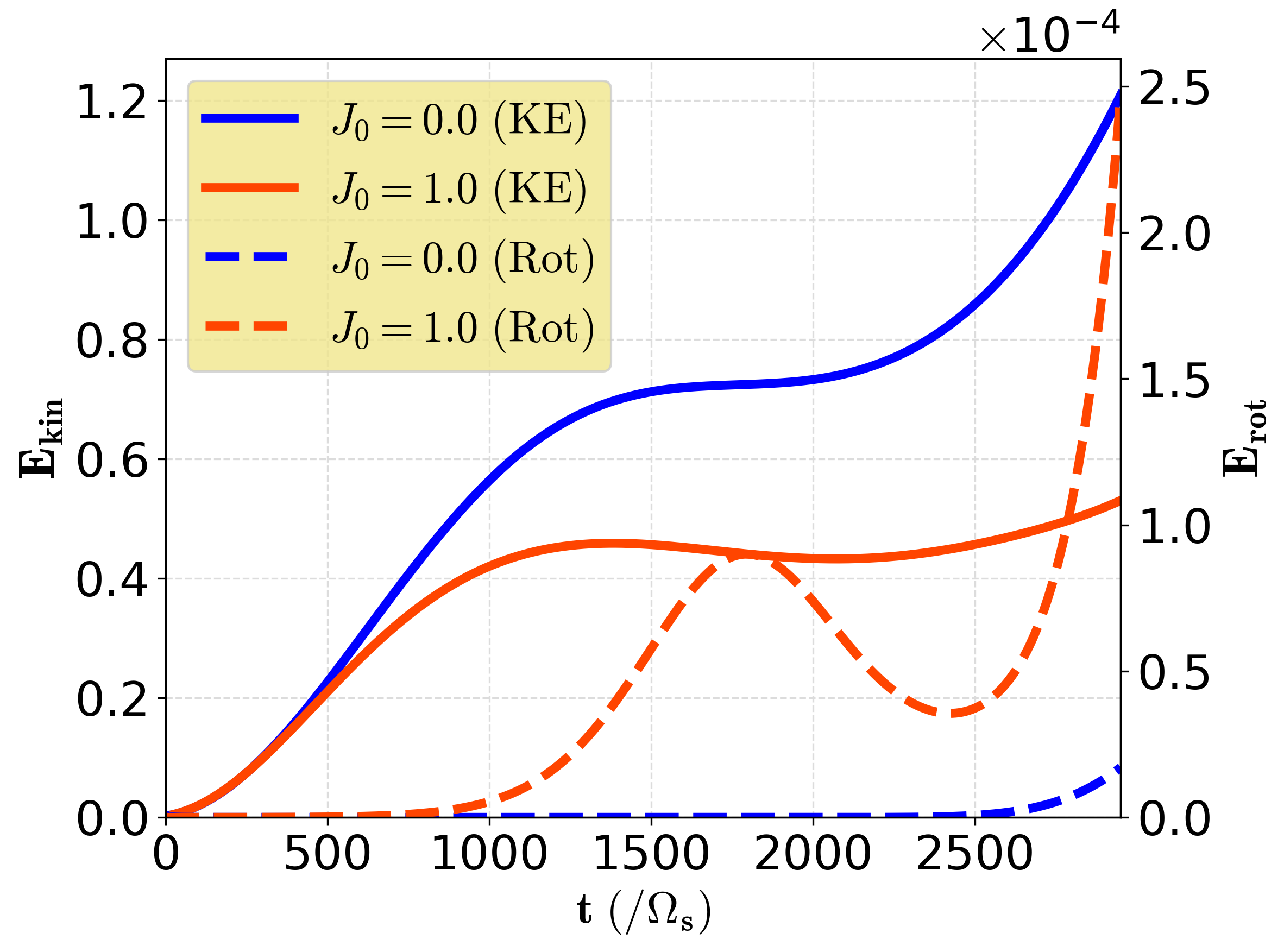}
    \caption{Temporal evolution of the translational kinetic energy and rotational energy for filaments with different current amplitudes. Increasing parallel current reduces the translational energy associated with radial propagation while enhancing rotational energy, indicating redistribution of filament energy into internal rotational motion.}
    \label{fig:energy}
\end{figure}

The change in translational kinetic and rotational energies also provides additional information about the filament dynamics, as shown in Fig.~\ref{fig:energy} where solid and dashed lines represent kinetic and rotational energies, respectively. Both quantities are computed from the $E\times B$ velocity field to quantify the redistribution of energy during filament evolution. The total kinetic energy is defined as
\begin{equation}
E_{\mathrm{kin}}=\frac{1}{2}\int\left(v_x^2+v_y^2\right)\,dx\,dy,
\end{equation}
while the rotational energy is calculated as
\begin{equation}
E_{\mathrm{rot}}=\frac{L^2}{2I},
\end{equation}
where $L$ and $I$ denote the angular momentum and moment of inertia evaluated in the center-of-mass frame of the filament.

At the early stage ($t \lesssim 500/\Omega_s$), both dipolar and current-carrying filaments show similar growth of the translational energy, while the rotational energy is weak, indicating that the dynamics are initially dominated by curvature-driven outward motion. When the nonlinear phase develops ($t \sim 50\text{-}150/\Omega_s$), clear differences are observed. For the dipolar filament ($J_0=0$), the translational kinetic energy continues to increase and dominates, consistent with stronger radial propagation, while the rotational energy remains relatively small. In contrast, the current-carrying filament ($J_0=1$) exhibits weaker growth of translational energy together with a significant increase in rotational energy.  This enhancement takes place over the same time interval that the spiral vorticity and angular momentum are developed, suggesting that some of the energy associated with the outward propagation is redistributed into the internal rotational motion. Such energy transfer provides a physical explanation for the suppressed radial velocity observed in current-carrying filaments.

%=========================================================================================
\subsection{Formation of Localized Shear Layers and Filament Velocity Reduction}
%=========================================================================================

\begin{figure*}
    \centering
    \includegraphics[width=0.45\linewidth]{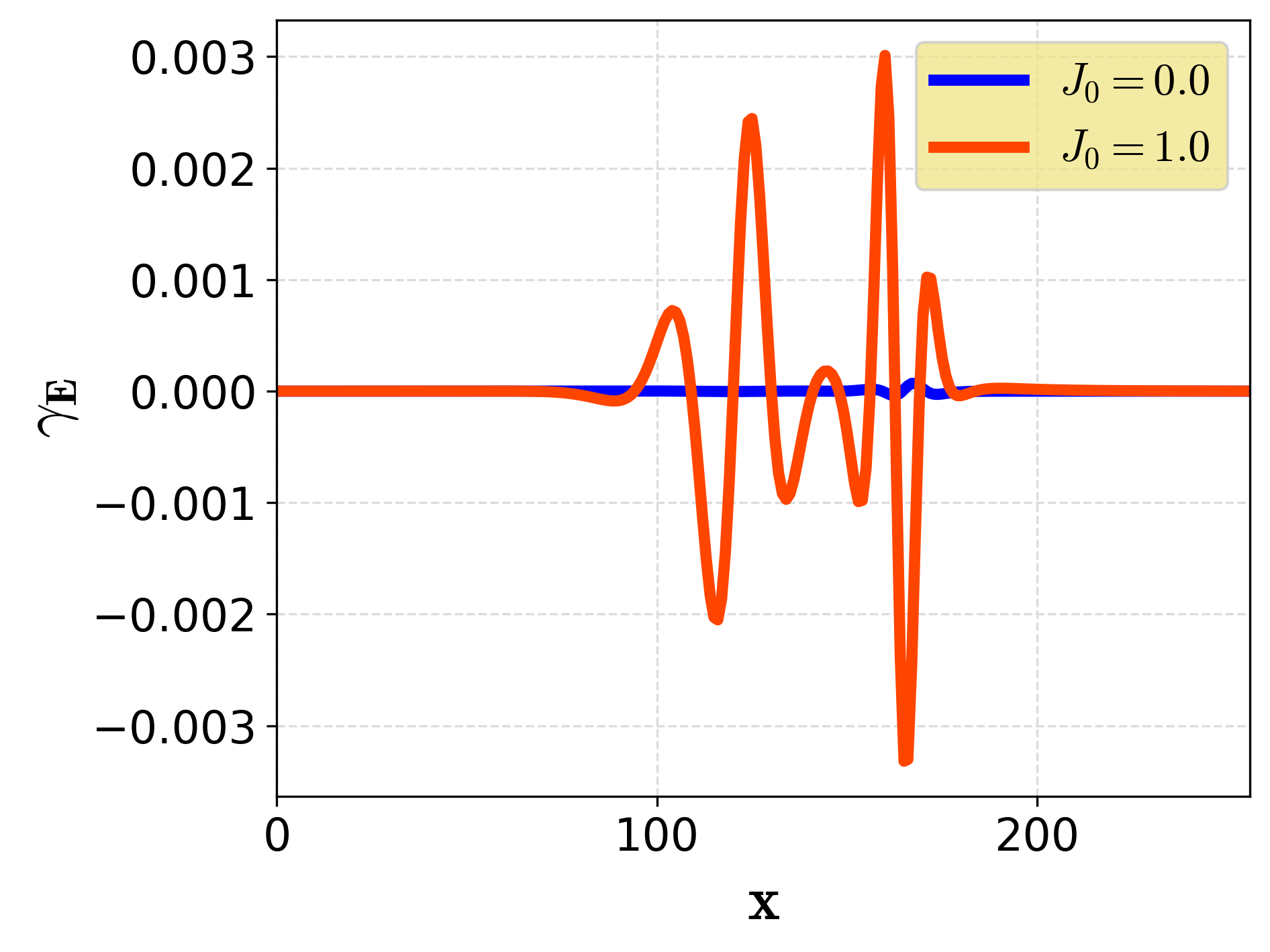}
    \includegraphics[width=0.45\linewidth]{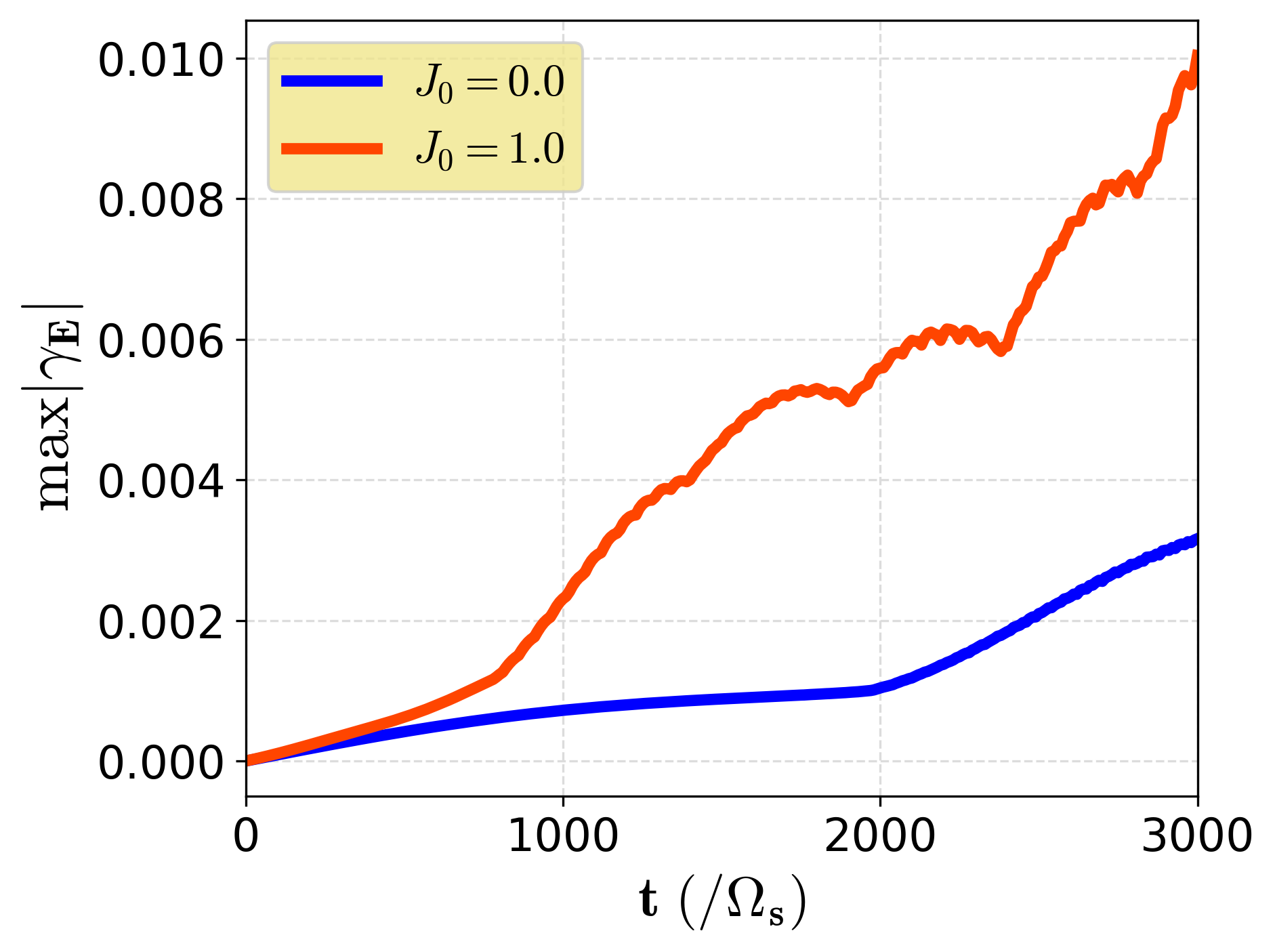}
    \caption{
    (Left) Radial shear profile at $t=1500/\Omega_s$ for the current-free ($J_0=0$) and current-carrying ($J_0=1$) filaments. The current-carrying filament develops significantly stronger localized regions of alternating positive and negative shear near the filament core, indicating enhanced velocity gradients associated with rotational motion. (Right) Time evolution of the maximum radial shear for both cases. Although the shear grows gradually in both filaments during the early stage, the current-carrying filament develops substantially larger shear during the nonlinear phase, consistent with the emergence of spiral vorticity and enhanced rotational dynamics.}
    \label{fig:shear}
\end{figure*}

The enhanced rotational motion generated by current-driven vorticity also leads to the formation of localized velocity shear. To quantify this effect, we calculate the radial shear of the poloidal \(E\times B\) flow, defined as the radial gradient of the poloidal velocity, \begin{equation} S_r = \frac{\partial v_y}{\partial x} = -\frac{\partial^2 \phi}{\partial x^2}, \end{equation} where \(v_y=-\partial \phi/\partial x\) is the normalized poloidal \(E\times B\) velocity.

Figure~\ref{fig:shear} shows the radial shear (left plot) profile at $t=1500/\Omega_s$ together with the temporal evolution of the maximum shear (right plot) for the current-free ($J_0=0$) and current-carrying ($J_0=1$) filaments. At $t=1500/\Omega_s$ the dipolar filament shows relatively weak shear localized near the center of the filament. In contrast, the current-carrying filament develops much stronger regions of alternating positive and negative shear around the filament core, indicative of the development of strong localized velocity gradients associated with rotational flow.

The difference is clearer in the time evolution of the maximum shear shown in Figure~\ref{fig:shear} (right). In the early stage, there is a slow growth of shear for both filaments as the filament evolves nonlinearly. However, in late times the maximal shear in the current-carrying filament grows much faster and reaches much larger values than in the dipolar case. This rise occurs during the same time as the spiral vorticity and rotational energy increase, suggesting that the enhanced shear is due to current-driven rotational dynamics. Strong local shear layers can distort the coherent outward motion and reduce the radial propagation. This provides an additional mechanism for suppression of velocity in current-carrying filaments. Such shear suppression of radial velocity has been widely discussed in relation to edge plasma turbulence and confinement studies~ \cite{Boedo_2002, burrell_1997}.

\begin{figure}
    \centering
    \includegraphics[width=0.99\linewidth]{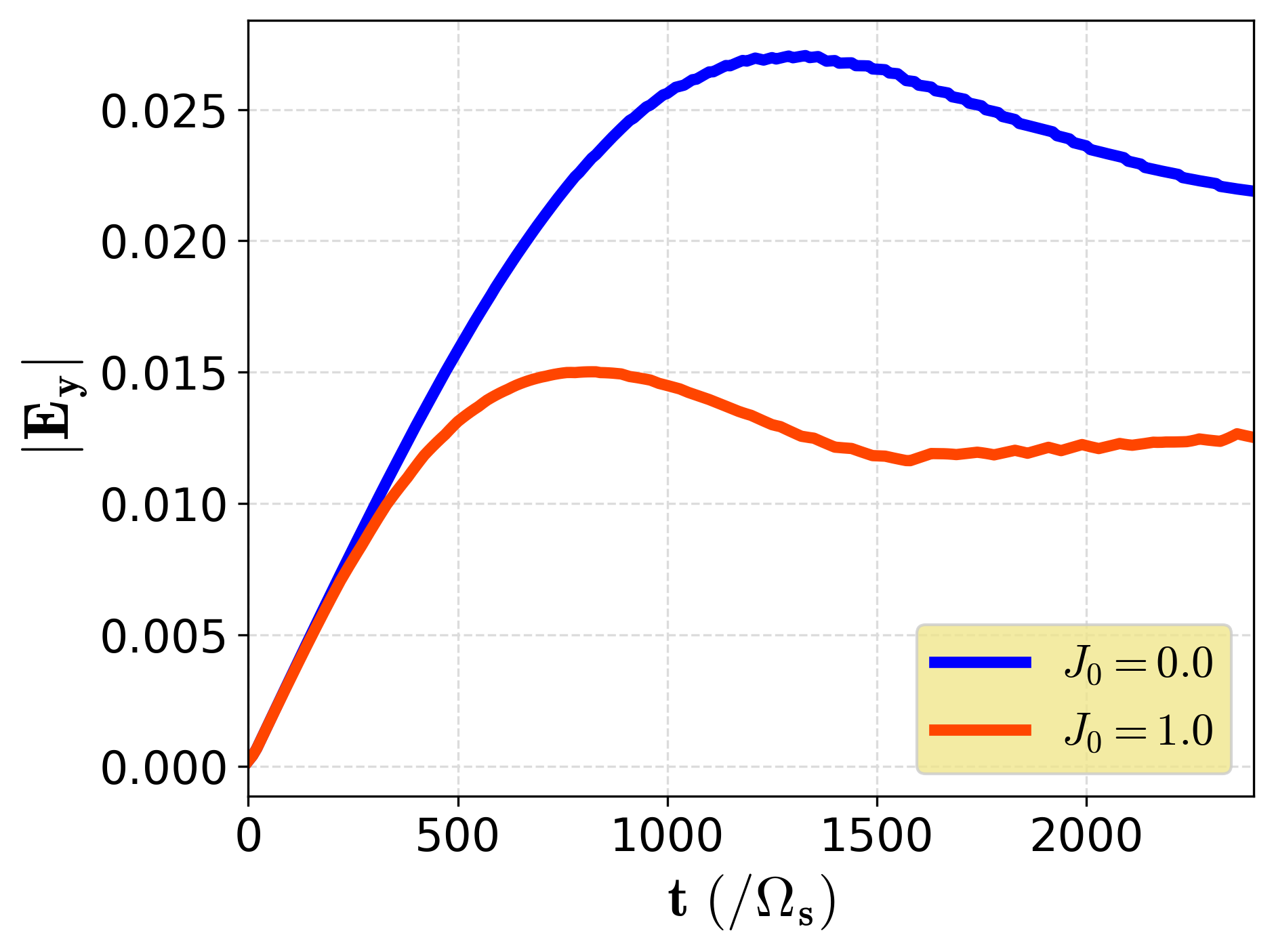}
    \caption{Time evolution of the poloidal electric field \(E_y\) for the current-free and current-carrying filaments. The current-carrying filament develops a weaker electric field structure compared to the dipolar filament, leading to reduced radial \(E\times B\) propagation.}
    \label{fig:electric_field}
\end{figure}

The reduction in radial velocity is closely connected to the changes in the poloidal electric field generated during the nonlinear evolution of the filament. Figure~\ref{fig:electric_field} presents the temporal evolution of the peak poloidal electric field, $E_y$, calculated at the filament center as $E_y=-\partial \phi/\partial y$. At the early stage of evolution, both the current-free ($J_0=0$) and current-carrying ($J_0=1$) filaments exhibit a gradual increase in the electric field due to the polarization of the density perturbation and the formation of the characteristic dipolar potential structure. However, clear differences emerge as the nonlinear phase develops. The current-free filament develops a significantly stronger poloidal electric field compared to the current-carrying filament, consistent with its stronger outward propagation. In contrast, the filament carrying finite parallel current exhibits a substantially weaker $E_y$ throughout the nonlinear stage. This reduction occurs during the same period in which current-driven vorticity becomes dominant, spiral vorticity emerges, and enhanced rotational motion, together with localized shear, develops.

The poloidal electric field directly regulates the radial $E\times B$ drift velocity, and consequently the weaker $E_y$ naturally produces a slower outward propagation \cite{Myra2006, dippolito_convective_2011}. These results indicate that the current-driven rotational self-organization of the filament redistributes energy away from the coherent radial motion and reduces the electric field responsible for transport. So the filament evolution is a sequence of parallel current driving vorticity generation, rotational dynamics and localized shear enhancement, effective poloidal electric field reduction and eventually radial velocity suppression.

%==========================================================================================================%

\section{Discussion and Conclusions}

%==========================================================================================================%

In this paper, we study the nonlinear dynamics of isolated current-carrying ELM filaments, using a reduced electromagnetic fluid model in slab geometry. It is observed that the unidirectional parallel current significantly changes the filament evolution by inhibiting the radial expansion and decreasing the outward transport. The filaments carrying current are more localized and propagate at a lower rate than the usual dipolar filaments, which grow predominantly due to the interchange motion driven by the curvature and form the characteristic mushroom-like shape~\cite{KRASHENINNIKOV2001368, dippolito_convective_2011}. The decrease in radial velocity suggests a weakening of the effective interchange drive by parallel current, which may in turn influence the transient particle and heat transport in tokamak edge plasmas~\cite{Zweben_2007, Kirk_2006}.

The main result of the present work is that the parallel current is an efficient electromagnetic source of vorticity generation. The conventional filament maintains a dipolar vorticity structure, whereas the current-carrying filament shows a pronounced spiral vorticity pattern in the nonlinear regime. The electromagnetic source term is very much larger than the usual curvature contribution, and directly imprints its topology on the vorticity field that is evolving. This vorticity reorganization is accompanied by enhanced angular momentum, increased rotational energy, and the formation of localized shear layers, indicative of a transition from conventional interchange-dominated propagation to a rotationally organized filament state. This shear generation may also further reduce transport by weakening coherent radial motion~\cite{burrell_1997, Boedo_2002, Wagner_2007}.

The present study highlights the important role of electromagnetic current effects in filament dynamics, particularly under high-\(\beta\) edge plasma conditions where ELM filaments are known to carry substantial field-aligned current~\cite{Kirk_2006, DIIID_ELM_PRL, myra_current_carrying_filament_2007}. Although the present model employs slab geometry and the cold-ion approximation, the identified mechanism of current-driven vorticity generation is expected to remain relevant in realistic tokamak plasmas. Future studies including toroidal geometry, finite ion temperature effects, and filament interactions will help assess the robustness of the observed rotational dynamics under experimentally relevant conditions.

%==========================================================================================================%

\begin{acknowledgments}

%==========================================================================================================%

The numerical simulations presented in this work were carried out using the Antya high-performance computing cluster at the Institute for Plasma Research (IPR).  A.S. acknowledges the Indian National Science Academy for the INSA Honorary Scientist position.
 
\end{acknowledgments}

\bibliographystyle{unsrt}
\bibliography{citation}

\end{document}